# Astrodynamical middle-frequency interferometric gravitational wave observatory AMIGO: Mission concept and orbit design

Wei-Tou Ni

*National Astronomical Observatories, Chinese Academy of Sciences, Beijing, China*
*State Key Laboratory of Magnetic Resonance and Atomic and Molecular Physics,*
*Wuhan Institute of Physics and Mathematics, Chinese Academy of Sciences, Wuhan 430071, China*
*Department of Physics, National Tsing Hua University, Hsinchu, Taiwan, 30013, ROC*
*weitou@gmail.com*

Gang Wang

*State Key Laboratory of Magnetic Resonance and Atomic and Molecular Physics,*
*Wuhan Institute of Physics and Mathematics, Chinese Academy of Sciences, Wuhan 430071, China*
*Shanghai Astronomical Observatory, Chinese Academy of Sciences, Shanghai, 200030, China*
*INFN, Gran Sasso Science Institute, I-67100 L'Aquila, Italy*
*INFN, Sezione di Pisa, Edificio C, Largo Bruno Pontecorvo, 3, I-56127, Pisa, Italy*
*gwanggw@gmail.com*

An-Ming Wu

*National Space Organization (NSPO), 8F, 9 Prosperity 1st Road, Science Park*
*Hsinchu, Taiwan, 30078, ROC*
*amwu@nspo.narl.org.tw*

AMIGO is a first-generation Astrodynamical Middle-frequency Interferometric GW Observatory. The scientific goals of AMIGO are: to bridge the spectra gap between first-generation high-frequency and low-frequency GW sensitivities; to detect intermediate mass BH coalescence; to detect inspiral phase and predict time of binary black hole coalescence together with neutron star coalescence for ground interferometers; to detect compact binary inspirals for studying stellar evolution and galactic population. The mission concept is to use time delay interferometry for a nearly triangular formation of 3 drag-free spacecraft with nominal arm length 10,000 km, emitting laser power 2-10 W and telescope diameter 300-360 mm. The design GW sensitivity in the middle frequency band is $3 \times 10^{-21}$ Hz$^{-1}$. Four options of orbits are under study: (i) Earth-like solar orbits (2-20 degrees behind the Earth); (ii) 600,000 km high orbit formation around the Earth; (iii) 50,000 km-250,000 high orbit formation around the Earth; (iv) near Earth-Moon L4 (or L5) halo orbit formation. All four options have LISA-like formations, that is the triangular formation is 60º inclined to the orbit plane. For AMIGO, the first-generation time delay interferometry is good enough for the laser frequency noise suppression. We also investigate for each options of orbits under study, whether constant equal-arm implementation is feasible. For the solar-orbit option, the acceleration to maintain the formation can be designed to be less than 15 nm/s$^2$ with the thruster requirement in the 15 μN range. AMIGO would be a good place to implement the constant equal-arm option. Fuel requirement, thruster noise requirement and test mass acceleration actuation requirement are briefly considered. From the orbit study, the solar orbit option is the first mission orbit choice. We study the deployment for this orbit option. A last-stage launch from 300 km LEO (Low Earth Orbit) to an appropriate 2-degree-behind-the-Earth AMIGO formation in 95 days requires only a Δv of 75 m/s.

## 1. Introduction

With the direct detection of the binary black hole mergers [1, 2, 3] and neutron star coalescence [4] by LIGO-Virgo collaboration, we have been fully ushered into the age of Gravitational Wave (GW) astronomy. Detection efforts over all GW frequency bands from cosmological frequency band (1 aHz−10 fHz) to ultra-high frequency band (over 1 THz) have been vigorously exerted (See, e.g. [5]). We have plotted the GW detector sensitivities and GW source strengths on single diagrams with ordinates showing characteristic strain, strain power spectral density (psd) amplitude and normalized GW spectral energy density respectively in 2015 [5]. A large part of GW detection efforts are concentrated on the GWs in the high frequency band 10 Hz–100 kHz (LIGO/VIRGO/KAGRA), the low frequency band 10 μHz–0.1 Hz (LISA) and the very low frequency band 300 pHz –100 nHz (PTAs). Fig. 1 adapted from Fig. 2 of Ref. [6] shows the strain psd (power spectral density) amplitude vs. frequency for various detectors and sources from 100 pHz to 10 kHz. For detailed explanation of the plot, see [5-8]. As can be seen from Fig. 1, there are 2 regions which are poor in the near-future projected sensitivities: (i) the middle frequency band, and (ii) the lower part (100 nHz–10 μHz) of the low frequency band. To possibly increase the sensitivity in the frequency band 0.1-10 μHz, Super-ASTROD with arm length of 9 AU has been proposed [9]. To have significant sensitivity in the frequency band 0.1−10 Hz and yet to be a first-generation candidate for space GW missions, we propose a middle-frequency GW mission AMIGO (Astrodynamical Middle-frequency Interferometric GW Observatory) with arm length 10,000 km. We have presented the basic mission concept in [6]. In Section 2, we discuss and elaborate on the mission concept. In Section 3 we treat both the solar orbit and earth orbit options mentioned in [6], and discuss deployment strategies. In Section 4, we calculate the first-generation TDIs (Time Delay Interferometry's) for each of orbit options studied in Section 3 and find that they all satisfy the frequency-noise suppression requirement. In Section 5, we investigate for each options of orbits under study, whether constant equal-arm implementation is feasible. For the solar-orbit option, the acceleration to maintain the formation can be designed to be less than 15 nm/s$^2$ with the thruster requirement in the 15 μN range. AMIGO would be a good place to implement the constant equal-arm option. Fuel requirement, thruster noise requirement and test mass acceleration actuation requirement are briefly considered. In Section 5, we present some discussions and give an outlook.

## 2. Mission Concept

A discussion of ground-based GW detector concepts to extend the present ground-based interferometers detection spectral range, i.e., the high-frequency GW band 10 Hz−100 kHz to middle-frequency band 0.1–10 Hz together with the plethora of potential astrophysical sources in this band is given in Harms et al. [10]. Harms et al. examine the potential sensitivity of three detection concepts (atom interferometers, torsion bar antennas and Michelson interferometers), estimate for their event rates and thereby, the sensitivity requirements for these detectors. They find that the scientific payoff from measuring astrophysical gravitational waves in this frequency band is great. However, although they find no fundamental limits to the detector sensitivity in this band, the remaining technical limits will be extremely challenging to overcome [10, 11]. SOGRO (Superconducting Omni-directional Gravitational Radiation Observatory) is another middle-frequency Earth-based GW detector concept [12-14]. The Newtonian-noise cancellation from infrasound and seismic surface fields is still very challenging [15].

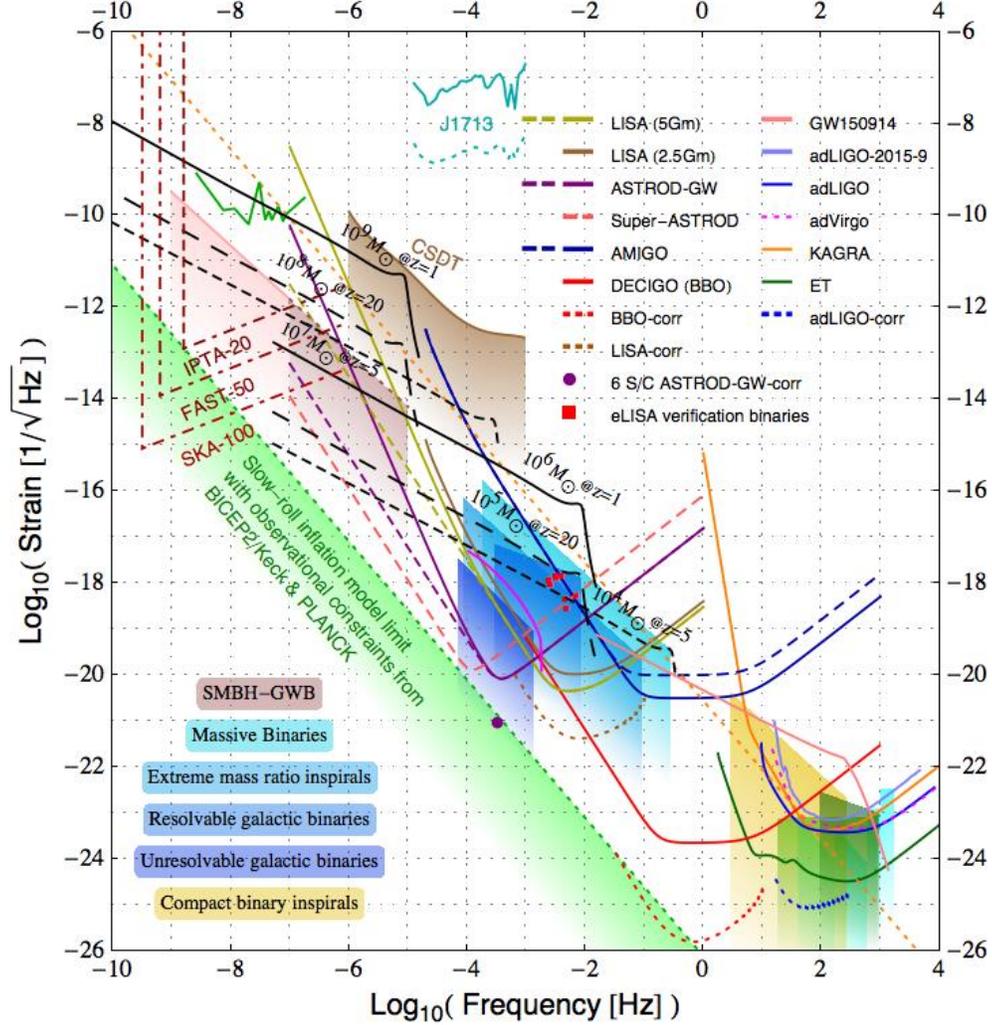

**Fig. 1.** Strain power spectral density (psd) amplitude vs. frequency for various GW detectors and GW sources. The black lines show the inspiral, coalescence and oscillation phases of GW emission from various equal-mass black-hole binary mergers in circular orbits at various redshift: solid line, $z = 1$; dashed line, $z = 5$; long-dashed line $z = 20$. See text of [7] for more explanation. The strain psd amplitude of GW150914 is calculated from its characteristic amplitude in Figure 1 of [8] using standard formula. The AMIGO design sensitivity is in solid blue while AMIGO baseline sensitivity is in dashed blue. The two curves merge together at lower frequency in the figure. [CSDT: Cassini Spacecraft Doppler Tracking; SMBH-GWB: Supermassive Black Hole-GW Background.]

## 2. 1. Basic concept

Astrodynamical Middle-frequency Interferometric GW Observatory is a *first-generation* middle-frequency mission concept with the following specification:

Arm length: 10,000 km (or a few times of this)
Laser power: 2 – 10 W
Acceleration noise: $S_a(f) = 9 \times 10^{-30} [1 + (10^{-4} \text{ Hz}/f)^2 + 16 (2 \times 10^{-5} \text{ Hz}/f)^{10}]$ m$^2$ s$^{-4}$ Hz$^{-1}$ (LPF has already achieved)
Orbits and formation: 4 options (all LISA-like formations):
    (i) Earth-like solar orbits (2-20 degrees behind the Earth orbit)
    (ii) 600,000 km geometric orbit formation
    (iii) 50,000-250,000 km geocentric orbit formation
    (iv) near Earth-Moon L4 (or L5) halo orbit formation
(From the study of this paper, the first option is the first choice.)

## 2. 2. Scientific goals

The scientific goals of AMIGO are:
(i)    to bridge the spectral gap between high-frequency and first-generation low-frequency GW sensitivities; Detecting intermediate mass BH coalescence;
(ii)    to detect inspiral phase and predict time of stellar-mass binary black hole coalescence together with neutron star coalescence or neutron star-black hole coalescence for ground interferometers, e.g., the inspiral GWs from sources like GW150914 as shown on Fig. 1;
(iii)    to detect compact binary inspirals for studying stellar evolution and galactic population.

## 2. 3. Sensitivity

To set the sensitivity level of AMIGO, we look into how the new LISA mission with arm length 2.5 Gm reach its design sensitivity.
  A new LISA proposal (Amaro-Seoane et al. [16]) was submitted to ESA on January 13th in response to the call for missions for the L3 slot in the Cosmic Vision Programme. On 20 June 2017, ESA announced the news that "The LISA trio of satellites to detect gravitational waves from space has been selected as the third large-class mission in ESA's Science programme (ESA 2017)." The basic concept is the same as the original LISA, but with arm length down-scaled to 2.5 Gm from 5 Gm. To distinguish this selected mission proposal from the original one or the NGO/eLISA, we call it LISA (2.5 Gm) or new LISA in case of possible ambiguity. Quoting from the proposal [16]:

"The observatory will be based on three arms with six active laser links, between three identical spacecraft in a triangular formation separated by 2.5 million km. Continuously operating heterodyne laser interferometers measure with pm Hz$^{-1/2}$ sensitivity in both directions along each arm, using well-stabilized lasers at 1064 nm delivering 2 W of power to the optical system. Using technology proven in LISA Pathfinder, the Interferometry Measurement System is using optical benches in each spacecraft constructed from an ultra-low expansion glass-ceramic to minimize optical path length changes due to temperature fluctuations. 30 cm telescopes transmit and receive the laser light to and from the other spacecraft. Three independent interferometric combinations of the light travel time between the test masses are possible, allowing, in data processing on the ground, the synthesis of two virtual Michelson interferometers plus a third null-stream, or "Sagnac" configuration." These two virtual Michelson interferometers are two TDIs. They could be two out of three TDI configurations X, Y and Z if they satisfy the noise requirement.

The new LISA design sensitivity is in [16, 17]. A simple analytical approximation of the design sensitivity is in Petiteau et al. [17] and used by Cornish and Robson [18]:

$$S_{Ln}^{1/2}(f) = (20/3)^{1/2} (1/L_L) \times [(1 + (f/(1.29 f_L))^2)]^{1/2} \times [(S_{Lp} + 4 S_a/(2\pi f)^4)]^{1/2} \ \text{Hz}^{-1/2}, \quad (1)$$

over the frequency range 20 μHz $< f <$ 1 Hz. Here $L_L$ = 2.5 Gm is the LISA arm length, $f_L = c/(2\pi L_L)$ is the LISA arm transfer frequency, $S_{Lp} = 8.9 \times 10^{-23}$ m$^2$ Hz$^{-1}$ is the white position noise, and

$$S_a(f) = 9 \times 10^{-30} \ [1 + (10^{-4} \text{ Hz}/f)^2 + 16 (2 \times 10^{-5} \text{ Hz}/f)^{10}] \ \text{m}^2 \text{ s}^{-4} \text{ Hz}^{-1}, \quad (2)$$

is the colored acceleration noise level. This new LISA design sensitivity curve is shown in Fig. 1.

For AMIGO, our baseline on the noise psd (power spectral density) amplitude assuming 2 W laser power, 30 cm telescopes and same acceleration noise as new LISA is:

$$S_{AMIGOn}^{1/2}(f) = (20/3)^{1/2}(1/L_{AMIGO}) \times [(1+(f/(1.29 f_{AMIGO}))^2)]^{1/2} \times [(S_{AMIGOp}+4 S_a/(2\pi f)^4)]^{1/2} \text{Hz}^{-1/2}, \quad (3)$$

over the frequency range of 20 μHz $< f <$ 1 kHz. Here $L_{AMIGO}$ = 0.01 × 10$^9$ m is the AMIGO arm length, $f_{AMIGO} = c/(2\pi L_{AMIGO})$ is the AMIGO arm transfer frequency, $S_{AMIGOp} = 1.424 \times 10^{-28}$ m$^2$ Hz$^{-1}$ is the (white) position noise level due to laser shot noise which is 16 × 10$^{-6}$ (=0.004$^2$) times that for new LISA. $S_a(f)$ is the same colored acceleration noise level in (2). The AMIGO baseline sensitivity (3) is plotted as AMIGO dashed curve in Fig. 1.

Since power and lower shot noise is crucial in reach better sensitivity in middle part of the sensitivity curve, we can use either (i) 10 W laser power and 36 cm ϕ telescope; or (ii) 2.688 W and 50 cm ϕ telescope as our design values of the AMIGO mission concept to gain a factor of 10 [≈ (10/2) × (36/30)$^4$ or ≈ (2.688/2) × (50/30)$^2$] for shot noise (power) design sensitivity. The AMIGO design sensitivity (3) is plotted as AMIGO solid curve in Fig. 1 by using $S_{\text{AMIGOp}} = 0.1424 \times 10^{-28}$ m$^2$ Hz$^{-1}$.

In the National Institute of Metrology in Beijing, high-efficient single frequency 1064 nm nonplanar ring Nd:YAG laser has been achieved with 4.54 W output power using 7.6 W diode pumping input at 885 nm [19]. Using this scheme or other, a space qualified stable Nd:YAG laser with 2.688 W output power is reasonable to expect in the near future. A laboratory model of SiC telescope mirror is under design analysis and construction [20].

With the development of 1 m ϕ SiC telescope mirror and 10 W laser power, the shot noise amplitude [$S_{\text{AMIGOp}}^{1/2}$] for strain will be decreased by another factor of 7.7 from the AMIGO design. We name this enhanced design the enhanced AMIGO (goal) sensitivity.

In Fig. 1, the strain psd amplitude of GW150914 is calculated from its characteristic amplitude in Figure 1 of [8] using standard formula of conversion. AMIGO with either baseline sensitivity or design sensitivity would detect the inspiral phase of GW150914 and predict the coalescence time for the benefit of doing multi-messenger astronomy. However, the design sensitivity has better coverage in detecting the inspiral phase of neutron star coalescence events.

The numerical TDIs for AMIGO would be easier to design compared to new LISA due to AMIGO's shorter arm length. X, Y, Z TDI configurations are well suited for AMIGO. However, experimental requirement on TDI is more stringent and needs developments.

*Technological readiness*. The main technological requirements of GW detection in space are (i) drag-free requirement; and (ii) requirement of measuring relative distance variation or relative velocity variation. LISA Pathfinder (LPF) launched on 3 December 2015, has achieved not only the drag-free requirement goal of this technology demonstration mission, but also has completely met the more stringent LISA drag-free demand [16, 17, 21, 22]. In short, LISA Pathfinder has successfully demonstrated the first generation drag-free technology requirement for space detection of GWs.

The requirement of measuring relative distance variation or relative velocity variation is in terms of spectral strain sensitivity. For space GW detection, the first-generation requirement is around $10^{-20}$ Hz$^{-1/2}$ sensitivity for measurement of strain psd amplitude. For measurement using unequal-arm laser interferometry, the requirement on laser stabilization is similar. However, the present laser stabilization has not reached this kind of stability. One needs to match the two optical paths using Time Delay Interferometry (TDI) to lessen the stability requirement. For TDI configurations and their numerical simulations for various missions, see Tinto and Dhurandhar [23], Wang and Ni [24] and references therein. Experimental demonstration of TDI in laboratory for LISA is worked out in 2010-2012 (Vine et al. [25], Mirtyk et al. [26]).

In space, Michelson type interferometry invariably involve large distances. The laser power received at the far end of the optical link is weak. To continue the optical path as required by TDIs, one needs to amplify it. The way of amplification is to track the optical phase of the incoming weak light with the local laser oscillator by optical phase-locking. At National Tsing Hua University, 2 pW weak-light homodyne phase-locking with 0.2 mW local oscillator has been demonstrated (Liao et al. [27, 28]). In JPL (Jet Propulsion Laboratory), Dick et al. [29] have achieved offset phase locking of local oscillator to 40 fW incoming laser light. More recently, Gerberding et al. [30] and Francis et al. [31] have phase-locked and tracked a 3.5 pW weak light signal and a 30 fW weak light signal respectively at reduced cycle slipping rate. For LISA, 85 pW weak-light phase locking is required. For ASTROD-GW, 100 fW weak-light phase locking is required. Hence, the weak level of these weak-light power requirements has achieved. In the future, the frequency-tracking, modulation-demodulation and coding-decoding needs development to make it a mature technology. This is also important for deep space CW (Continuous Wave) optical communication.

*Arm length.* As shown in Fig. 1, typical frequency sensitivity spectrum of strain psd amplitude for space GW detection consists of three regions, the acceleration/local gravity gradient/vibration noise dominated region, the shot noise (flat for current space detector projects like LISA in strain psd) dominated region, if any, and the antenna response restricted region. The detector sensitivity in the lower frequency region is constrained by vibration, acceleration noise or gravity-gradient noise. The detector sensitivity of the higher frequency part is constrained by antenna response (or storage time). In a power-limited design, sometimes there is a middle flat region in which the sensitivity is limited by the photon shot noise. [7, 32-34]

The shot noise sensitivity in the strain for GW detection is inversely proportional to $P^{1/2}L$ with $P$ the received power and $L$ the distance or arm length. Since $P$ is inversely proportional to $L^2$ and $P^{1/2}L$ is constant, this sensitivity limit is independent of the distance. For 1-2 W emitting power, the limit is around $10^{-20}$–$10^{-21}$ Hz$^{-1/2}$ (depending on telescope diameter/laser beam divergence). As noted in the LISA study [32], making the arms longer shifts the time-integrated sensitivity curve to lower frequencies while leaving the bottom of the curve at the same level. Hence, ASTROD-GW with longer arm length has better sensitivity at lower frequency. e-LISA, ALIA, TAIJI, and GW interferometers in Earth orbit have shorter arms and therefore have better sensitivities at higher frequency. The arm length of AMIGO is shorter than new LISA by 250 fold; hence the bottom flat sensitive region is shifted by 250 fold from LISA mHz frequency to middle frequency (0.1-10 Hz).

## 3. Orbit Design

In subsection 2.1, we have listed both heliocentric and geocentric options for the AMIGO orbit configuration. These options are for studying various purposes to be discussed in section 6. In subsection 3.1, we treat the following three heliocentric AMIGO orbit choices to illuminate on various orbit design possibilities and issues:

(i) AMIGO-S: AMIGO-Earth-like solar orbits with formation varying between 8 and 12 degrees behind the Earth orbit starting at epoch JD2462316.0 (2029-Jun-28th 12:00:00),

(ii) AMIGO-S: AMIGO-Earth-like solar orbits with formation varying between

2 and 6 degrees behind the Earth orbit starting at epoch JD2462416.0 (2029-Oct-6th 12:00:00),

(iii) AMIGO-S: AMIGO-Earth-like solar orbits with formation varying between 2 and 4 degrees behind the Earth orbit starting at epoch JD2462503.0 (2030-Jan-1st 12:00:00), in J2000 equatorial (Earth mean equator and equinox) solar-system-barycentric coordinate system.

In subsection 3.2, we explore two geocentric orbit possibilities. Specifically, we treat and discuss the following cases:

(i) AMIGO-E1: 100,000 km geocentric orbit formation,

(ii) AMIGO-EM: near Earth-Moon L4 halo orbit formation.

Other geocentric options (AMIGO-E2: 50,000 km geocentric, AMIGO-E3: 150,000 km geocentric, AMIGO-E4: 200,000 geocentric, AMIGO-E5: 600,000 geocentric, and AMIGO-EM: near Earth-Moon L5 halo orbit formation have been studies; similar conclusions cab be reached in Section 6. For the brevity of this paper (already long), we present them elsewhere.

## 3. 1. AMIGO-S mission orbit optimization and deployment

The goal of the AMIGO-S mission orbit optimization is to equalize the three arm lengths of the AMIGO-S formation as much as possible and to make the relative line-of-sight velocities smaller than 0.1 m/s between three pairs of spacecraft. To begin with, we follow the initial orbit selection algorithm discussed in Ref. 35, which was used in our previous works (Refs. 24, 36, and 37), and calculate at epoch JD2462503.0 (2030-Jan-1st 12:00:00) for 10°, 4° and 2° behind-the-Earth configurations. We evolve forward the configurations to find a suitable duration. The initial conditions are adjusted slightly according to our previous work (Refs. 24, 36, and 37) to find optimal duration. To extend the possible duration, we also evolve backward for the 10° and 4° cases until JD2462316.0 (2029-Jun-28th 12:00:00) and JD2462416.0 (2029-Oct-6th 12:00:00) respectively using CGC3.0 ephemeris framework. We use these epochs as starting time of the science observation and list the positions and velocities of S/C at these epochs in Table 1. The AMIGO spacecraft orbits are then calculated respectively for 600 days, 250 days and 80 days using CGC3.0. The variations of arm lengths, the Doppler velocities between the AMIGO S/Cs, the formation angles and the angle between Earth and barycenter of S/Cs are drawn in Fig. 2, Fig. 3 and Fig. 4 respectively. In Fig. 2, the variations of arm lengths are within ±0.6% and Doppler (relative) velocities in the line of sight direction are within ±0.035 m/s; the formation angles breathe less than ±0.6°; the angle between barycenter of S/C and Earth in 600 days starts at 10° behind Earth and varies between 8° and 12° with a quasi-period of variation about 1 sidereal year due mainly to Earth's elliptic motion. In Fig. 3, the variations of arm lengths are within ±1.7% and Doppler (relative) velocities in the line of sight direction are within ±0.09 m/s; the formation angles breathe less than ±1.5°; the angle between barycenter of S/C and Earth in 250 days starts at 2° behind Earth and varies between 2° and 6° with a quasi-period of variation about 1 sidereal year due mainly to Earth's elliptic motion. In Fig. 4, the variations of arm lengths are within ±1.7% and Doppler (relative) velocities in the line of sight direction are within ±0.15 m/s; the angle between barycenter of S/C and Earth in 80 days starts at 2° behind Earth and varies between 2° and 4°.

**Table 1.** Initial conditions of 3 S/C for 3 choices of AMIGO orbit formations with 10,000 km nominal arm length respectively at epoch JD2462316.0 (2029-Jun-28th 12:00:00), at epoch JD2462416.0 (2029-Oct-6th 12:00:00) and at epoch JD2462503.0 (2030-Jan-1st 12:00:00) in J2000 equatorial (Earth mean equator and equinox) solar-system-barycentric coordinate system.

| | | Position and Velocity of 3 S/C for AMIGO-S (8°-12° behind Earth) at Epoch JD2462316.0 (2029-Jun-28th 12:00:00) | Position and Velocity of 3 S/C for AMIGO-S (2°-6° behind Earth) at Epoch JD2462416.0 (2029-Oct-6th 12:00:00) | Position and Velocity of 3 S/C for AMIGO-S (2°-4° behind Earth) at Epoch JD2462503.0 (2030-Jan-1st 12:00:00) |
|---|---|---|---|---|
| S/C1 Position (AU) | X | -6.302152839887E-02 | 9.8142175696900E-01 | -1.505970243492E-01 |
| | Y | -9.152327833959E-01 | 1.74218816072566E-01 | 9.072202343045E-01 |
| | Z | -3.968044812826E-01 | 7.5571613241614E-02 | 3.934063541577E-01 |
| S/C1 Velocity (AU/day) | $V_x$ | 1.717237542522E-02 | -3.232782952641E-03 | -1.700730831376E-02 |
| | $V_y$ | -1.015996662090E-03 | 1.5492007968084E-02 | -2.381854275059E-03 |
| | $V_z$ | -4.405295611268E-04 | 6.7173218829668E-03 | -1.032610751502E-03 |
| S/C2 Position (AU) | X | -6.305233647362E-02 | 9.8141402230888E-01 | -1.506257047811E-01 |
| | Y | -9.152796107267E-01 | 1.7417831148714E-01 | 9.072093796258E-01 |
| | Z | -3.967678883765E-01 | 7.5519749833036E-02 | 3.933470061939E-01 |
| S/C2 Velocity (AU/day) | $V_x$ | 1.717184116187E-02 | -3.232406473962E-03 | -1.700770604481E-02 |
| | $V_y$ | -1.016019245713E-03 | 1.5492740749576E-02 | -2.382374955735E-03 |
| | $V_z$ | -4.410074255659E-04 | 6.7167092533236E-03 | -1.032293794868E-03 |
| S/C3 Position (AU) | X | -6.298507576262E-02 | 9.8144991412237E-01 | -1.505596242826E-01 |
| | Y | -9.152797382451E-01 | 1.7415689199138E-01 | 9.072186495149E-01 |
| | Z | -3.967728403737E-01 | 7.5572545721927E-02 | 3.933510251827E-01 |
| S/C3 Velocity (AU/day) | $V_x$ | 1.717191754514E-02 | -3.232312831532E-03 | -1.700780995129E-02 |
| | $V_y$ | -1.015881495969E-03 | 1.549225801561E-02 | -2.381459758927E-03 |
| | $V_z$ | -4.398650268940E-04 | 6.716454443883E-03 | -1.032982407703E-03 |

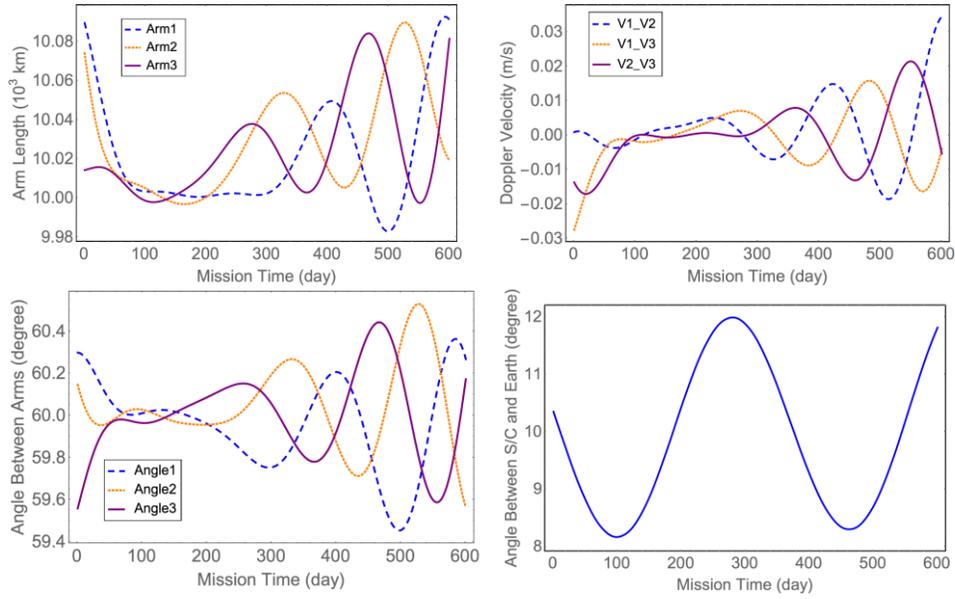

**Fig. 2.** Variations of the arm lengths, the Doppler (relative) velocities, the formation angles and the angle between barycentre of S/C and Earth in 600 days for the AMIGO-S (8°-12° behind Earth) 3-S/C formation with initial conditions given in column 3 of Table 1.

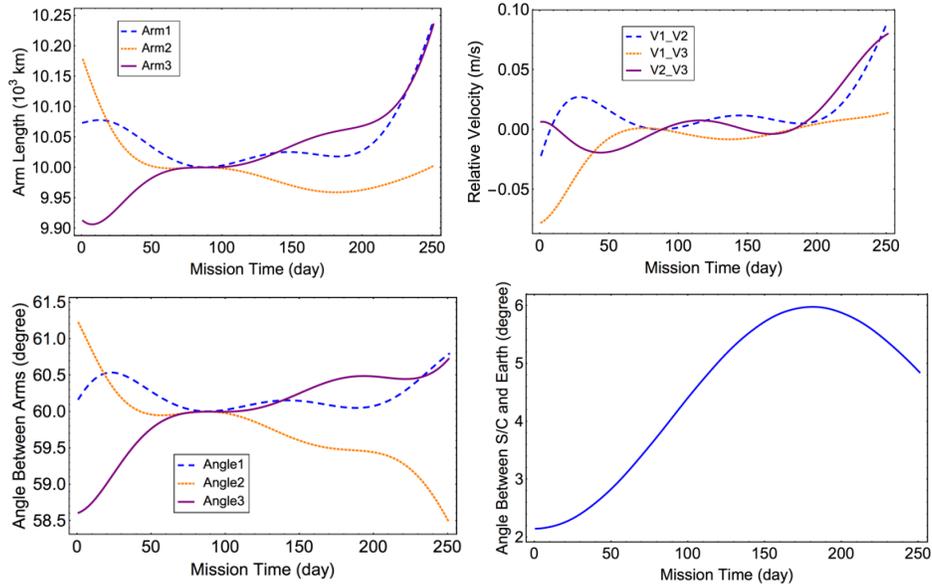

**Fig. 3**. Variations of the arm lengths, the Doppler (relative) velocities, the formation angles and the angle between barycenter of S/C and Earth in 250 days for the AMIGO-S (2°-6° behind Earth) 3-S/C formation with initial conditions given in column 4 of Table 1.

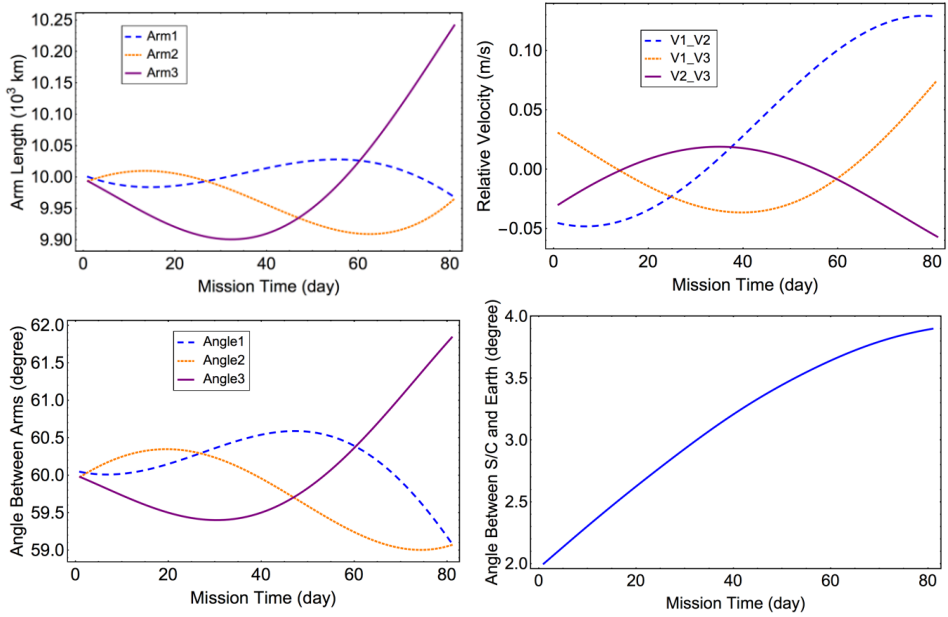

**Fig. 4**. Variations of the arm lengths, the Doppler (relative) velocities, the formation angles and the angle between barycenter of S/C and Earth in 80 days for the AMIGO-S (2°-4° behind Earth) 3-S/C configuration with initial conditions given in column 5 of Table 1.

*Deployment.* In the case of AMIGO-S orbit formation with 2°-4° trailing angle to Earth orbit, if we evolve the orbits back in time from Epoch JD2462503.0 (2030-Jan-1st 12:00:00), the formation would go near to Earth and through to the other side of Earth. If we launch at the appropriate near Earth point, this would need no S/C Δv to connect to the starting point at 2°, and so no Δv for deployment. From low Earth orbit, at appropriate time and appropriate launch velocity, the deployment Δv at Epoch JD2462503.0 (2030-Jan-1st 12:00:00) would be small. We list out findings in Table 1. The Δv's for the corresponding transfer times of 22.5 day to 180 day are shown in Table 2 and Fig. 5. From the table or figure, one can find that the minimum Δv for optimal deployment is below 0.075 km/s (75 m/s) and corresponding transfer time is around 95~100 day for the three spacecraft.

The deployment procedures are the direct transfer from the low Earth orbit (LEO) with 300 km altitude to the mission orbit [38]. JPL DE431 solar ephemerides [39] are utilized for the initial states of the solar system for the equations of motion of the celestial bodies and the spacecraft. At Epoch JD[2462503.0 − transfer time], an appropriate point in the 300 km altitude orbit is chosen for final-stage firing to obtain a final stage launch velocity. We employ Lambert fixed-end-positions-and-time method to find the transfer orbit solution instead of using initial position and initial velocity to find the trajectory solution. With this solution the 2 velocities at 2 end points (initial position and final position) are obtained. The initial velocity minus the LEO orbit velocity is the velocity required for the last stage launcher to supply. The science orbit starting velocity minus the final transfer orbit velocity is the velocity that the thrusters of the spacecraft need to supply; this is the Δv listed in Table 2. The detailed methods are explained in [38]. The detailed calculation for AMIGO will be presented in [40].

The deployment for other solar orbit options of AMIGO is discussed in [38, 40, 41]. The purpose of this discussion is to show that the deployment of mission formation to 2°-4° trailing angle to Earth orbit does not require much Δv and the method is simple. It would be a good place for deploying a pathfinder mission as well as a dedicated mission.

**Table 2**. Transfer time vs Δv needed for 3 S/C to go into the science orbit formation at Epoch JD24XXX462503.0 (2030-Jan-1st 12:00:00) to have the positions and velocities listed in the column 5 of Table 1.

| Transfer Time (day) | SC1 Δv (km/s) | SC2 Δv (km/s) | SC3 Δv (km/s) | Mean Δv (km/s) |
|---|---|---|---|---|
| 22.5 | 2.0360 | 2.0327 | 2.0381 | 2.0356 |
| 30.0 | 1.3341 | 1.3324 | 1.3362 | 1.3342 |
| 45.0 | 0.6415 | 0.6421 | 0.6438 | 0.6425 |
| 60.0 | 0.3255 | 0.3283 | 0.3285 | 0.3274 |
| 67.5 | 0.2288 | 0.2328 | 0.2321 | 0.2313 |
| 80.0 | 0.1235 | 0.1298 | 0.1275 | 0.1269 |
| 85.0 | 0.0980 | 0.1051 | 0.1022 | 0.1018 |
| 90.0 | 0.0819 | 0.0894 | 0.0861 | 0.0858 |
| 95.0 | 0.0750 | 0.0823 | 0.0788 | 0.0787 |
| 100.0 | 0.0755 | 0.0819 | 0.0786 | 0.0787 |
| 112.5 | 0.0912 | 0.0951 | 0.0928 | 0.0930 |
| 120.0 | 0.1022 | 0.1050 | 0.1032 | 0.1035 |
| 135.0 | 0.1188 | 0.1205 | 0.1193 | 0.1195 |
| 150.0 | 0.1284 | 0.1298 | 0.1288 | 0.1290 |
| 157.5 | 0.1312 | 0.1326 | 0.1317 | 0.1318 |
| 180.0 | 0.1365 | 0.1380 | 0.1367 | 0.1370 |

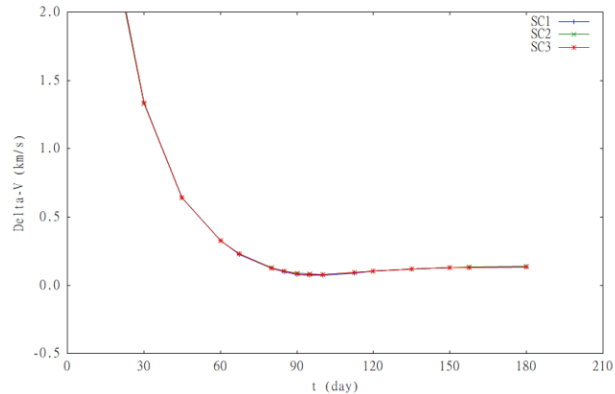

**Fig. 5**. Deployment delta-Vs (Δv's) of SC1, SC2, and SC3 for transfer times of 22.5-180.0 day.

## 3.2. AMIGO-E orbits

In this subsection, we explore two geocentric orbit possibilities. Specifically, we treat and discuss the following two cases:

(i) AMIGO-E1: 100,000 km geocentric orbit formation,

(ii) AMIGO-EM: near Earth-Moon L4 halo orbit formation.

### 3.2.1. AMIGO-E1 orbit

The construction and propagation of Earth-Moon (EM) LISA-like formation are described as follows:

(1) Generate solar ephemerides from the positions and velocities of Sun, Earth, Moon, 7 other major planets and Pluto from JPL DE431 [39] at a specific epoch.
(2) Compute the orbit elements of Moon to Earth.
(3) Specify the orbit elements of EM LISA-Like formation for semimajor axis (SMA) of 100,000 km and arm length of 10,000 km, and obtain inclination of 0.745 deg and eccentricity of 0.00751 (Fig.6).
(4) Propagate for 360 days.
(5) Output arm lengths, differences of arm lengths, Doppler velocities, and breathing angles.

According to the linearized orbit equation [42], a formation of regular triangle tilted at 60° with respect to the Moon's orbit plane can be constructed from Fig. 6 like LISA mission [16] and TAIJI mission [43].

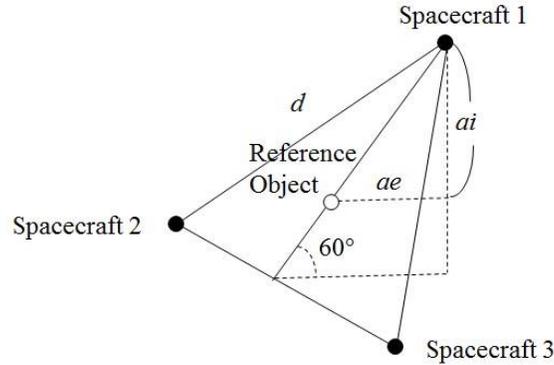

**Fig. 6**. Formation geometry from linearized orbit equation.

For a regular triangle, one has the following relations.

$$ae = d/(2\sqrt{3}), \; ai = d/2 \qquad (4)$$

Given the distance between the satellites $d$ and the semimajor axis of the orbits $a$, the eccentricity $e$ and the inclination $i$ can be obtained.

The orbit parameters of the three spacecraft in the ecliptic formation are selected as shown in Table 3, in which the semimajor axis $a$ is 100,000 km and the distance $d$ is 10,000 km. The three spacecraft have different in RAAN (right ascension of the ascending node) and mean anomaly only.

**Table 3.** Orbit parameters of AMIGO-E1 mission with arm length 10,000 km

| Element | Spacecraft 1 | Spacecraft 2 | Spacecraft 3 |
|---|---|---|---|
| Semimajor Axis | 100,000 km | 100,000 km | 100,000 km |
| Eccentricity | .00751 | .00751 | .00751 |
| Inclination | 2.864 deg | 2.864 deg | 2.864 deg |
| Right Ascension of the Ascending Node | 0 deg plus that of Moon plus 180 deg | 120 deg plus that of Moon plus 180 deg | 240 deg plus that of Moon plus 180 deg |
| Argument of Perihelion | -90 deg | -90 deg | -90 deg |
| Mean Anomaly | 180 deg | 60 deg | -60 deg |

Similar orbit parameters can be obtained for other formations with different arm lengths.

*Simulation of the Arm Lengths.* JPL DE431 solar ephemerides [39] are utilized for the initial states of the solar system for the equations of motion of the celestial bodies and the spacecraft. The propagation period is 360 days with time step of 0.0125 day for the simulations of arm lengths, arm length differences, breathing angle, and Doppler velocities. We also take into account of Sun's quadrupole and Earth oblateness. Due to perturbations, the relative positions of the three spacecraft will gradually deviate from the initial configuration, we propagate forward for 180 days and backward for 180 days to obtain nearly optimal orbits. The initial states of three spacecraft are given in Table 4.

Fig. 7 shows the 360-day evolution of AMIGO-E1 formation for AMIGO with Earth at the center of inertial frame of solar system. Fig. 8 shows the evolution of arm lengths, arm length differences, Doppler velocities, and formation breathing angles.

### 3.2.2. AMIGO-EM-L4

For the case of formation near EML4, the initial states of three spacecraft are obtained near Earth-Moon L4 by the same method as in section 3.2.1. and given in the last 3 rows of Table 4. Evolution of formation in rotating frame of Earth-Moon system is shown in Figure 9, and the simulation of arm lengths and others are shown in Figure 10.

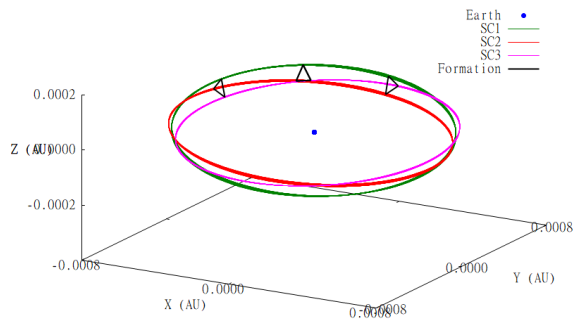

**Fig. 7.** Evolution of AMIGO-E1 formation in inertial frame of solar system with Earth at center for 360 days, in which the formation at three mission times are also plotted.

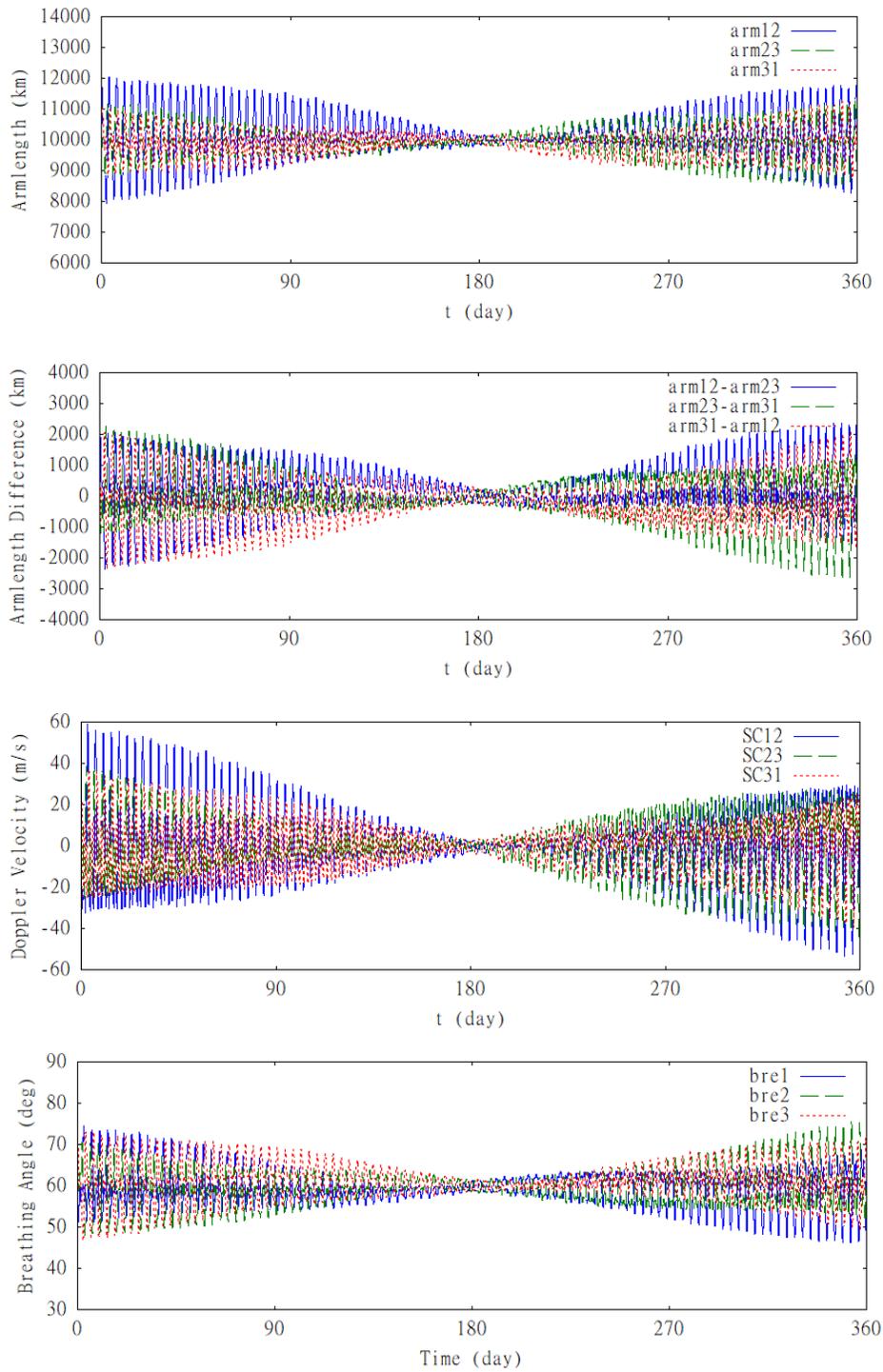

**Fig. 8**. Evolution of arm lengths (upper), arm length differences (middle upper), Doppler velocities (middle lower), breathing angles and (lower) for 360 days with SMA=100,000 km.

Table 4. The positions and velocities (barycentric t, x, y, z, u, v, w) of Sun, Earth, Moon, 7 other major planets and Pluto from JPL DE431 together with 3 S/C of AMIGO-E1 formation with SMA=100,000 km and AMIGO-EM-L4 formation near EML4 at Epoch=2025.3.17 16:57:09.7568 (180.000000000 day of the Fig. 8 and Fig. 10).

Sun:
   -0.005181332    -0.005152133    0.000170201    0.000007277    -0.000002737    -0.000000124

Earth:
   -0.998951054    0.046955159    0.000174869    -0.001176641    -0.017242277    0.000001377

Moon:
   -1.001188467    0.045429542    0.000023137    -0.000860108    -0.017702271    -0.000039635

Mercury:
   -0.317776491    0.152777674    0.041748071    0.008463371    -0.023927374    -0.000261137

Venus:
   -0.719230413    0.074241451    0.042460931    -0.002338412    -0.020194732    -0.000142128

Mars:
   -1.330534231    0.995471666    0.053640662    -0.007898552    -0.009976555    -0.000015266

Jupiter:
   0.479299271    5.079735791    -0.031791800    0.007599172    0.001069077    0.000165646

Uranus:
   10.849351373    16.242926717    -0.080228995    0.003299556    0.002001266    0.000050249

Saturn:
   9.500540049    -1.353697744    -0.354728923    0.000476629    0.005511223    -0.000114681

Naptune:
   29.876757360    -0.400071484    -0.680300318    0.000021723    0.003157448    -0.000065863

Pluto:
   18.432246127    -29.941508239    -2.127792280    0.002749050    0.000947810    -0.000889272

AMIGO-E1 SC1:
   -0.999335513    0.047517032    0.000272325    -0.002100765    -0.017875474    0.000006383

AMIGO-E1 SC2:
   -0.999350053    0.047477933    0.000220687    -0.002120535    -0.017930002    0.000055000

AMIGO-E1 SC3:
   -0.999294184    0.047516214    0.000220384    -0.002159110    -0.017874470    -0.000041811

AMIGO-EM-L4 SC1:
   -0.998755885    0.044379094    0.000007771    -0.000595460    -0.017196356    -0.000027752

AMIGO-EM-L4 SC2:
   -0.998727423    0.044414121    -0.000041378    -0.000588865    -0.017192515    -0.000021137

AMIGO-EM-L4 SC3:
   -0.998794224    0.044408843    -0.000038030    -0.000589035    -0.017199169    -0.000035020

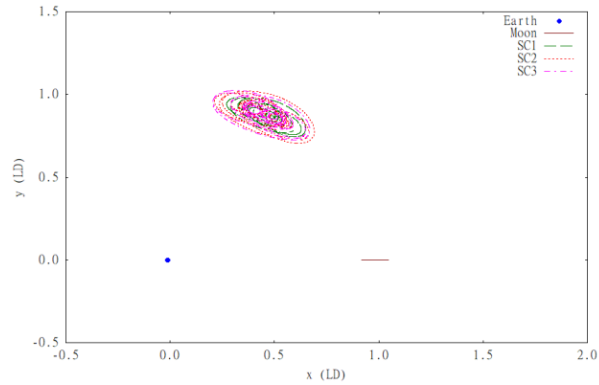

Figure 9. Evolution of the formation near EML4 in rotating frame of Earth-Moon system for 360 day.

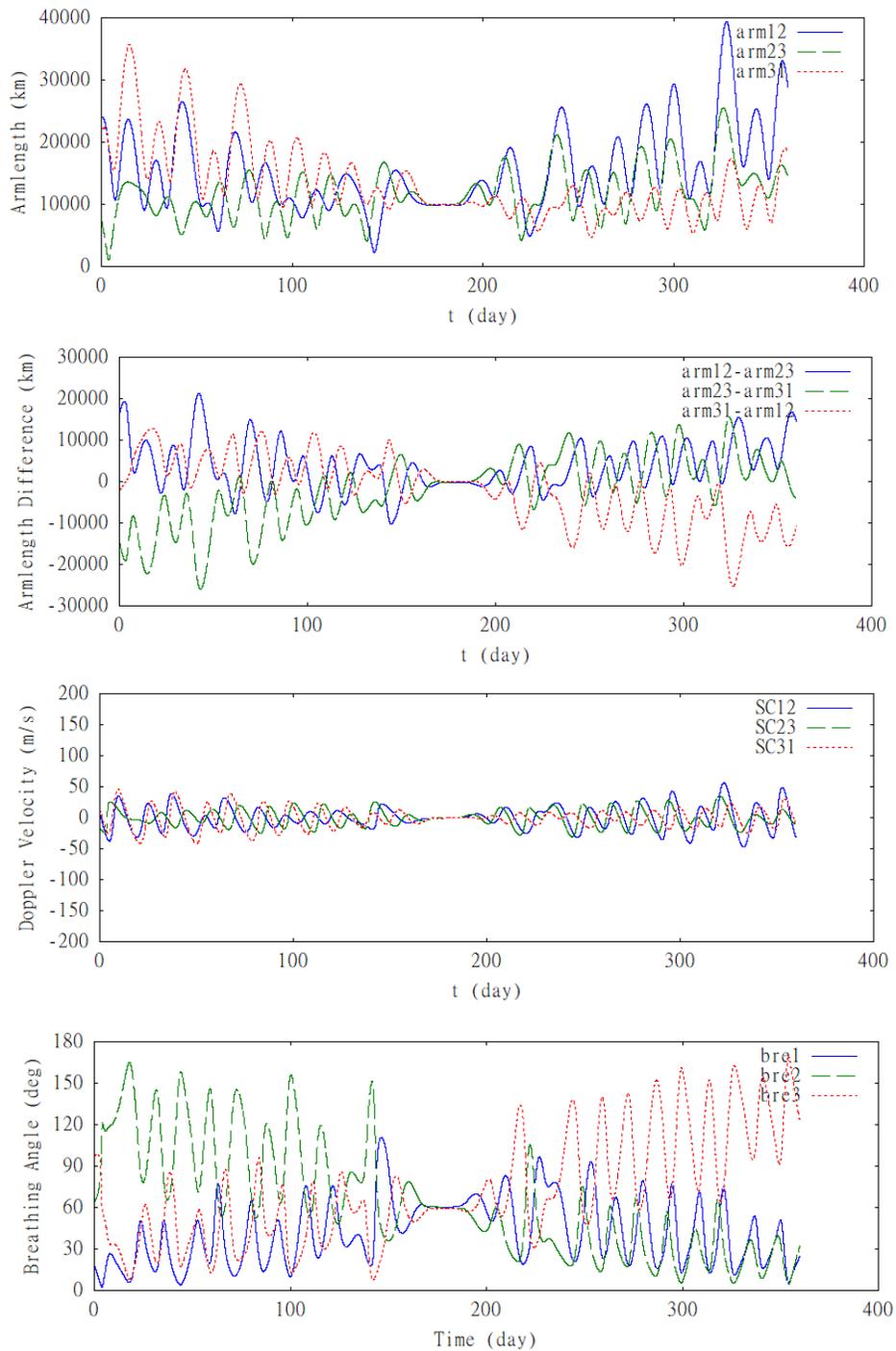

**Figure 10**. Evolution of arm lengths (upper), arm length differences (middle upper), breathing angles (middle lower), and Doppler velocities (lower) for 360 days for formation near EML4.

## 4. Time Delay Interferometry

In this section, we calculate the difference between the two path lengths for TDI configurations and plot the difference as function of the signal arriving epoch of TDI in the first-generation TDI configurations --- Michelson X, Y & Z; Sagnac $\alpha$, $\beta$ & $\gamma$; Relay U, V & W; Beacon P, Q & R; Monitor E, F & G for various orbit choices of section 3 for AMIGO-S, AMIGO-E1, and AMIGO-EM. We use the iteration and interpolation methods to calculate the time in the barycentric coordinate system as in our early papers [24]. We do this for AMIGO-S-8-12deg in section 4.1, for AMIGO-S-2-6deg in section 4.2, for AMIGO-S-2-4deg in section 4.3, for AMIGO-E1 in section 4.4, and for AMIGO-EM-L4 in section 4.5. In section 4.6, we summarize the results in Table 6.

### 4.1. Numerical results of the first-generation TDI for AMIGO-S-8-12deg

*4.1.1. Unequal-arm Michelson* X*,* Y *&* Z *TDI*s *and their sum* X+Y+Z *for AMIGO-S-8-12deg* (Fig. 11)

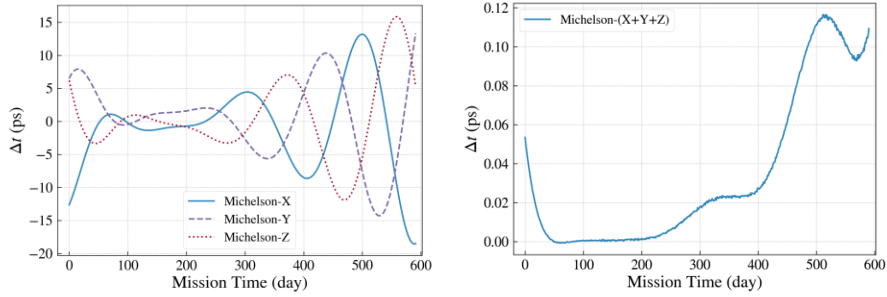

**Fig. 11.** The optical path length differences for Unequal-arm Michelson X, Y, & Z TDIs (left panel) and their sum X + Y + Z (right panel) for AMIGO-S-8-12deg. There is a clear cancellation of optical path length differences by 2 orders of magnitudes in the sum.

*4.1.2. Sagnac α, β & γ; Relay* U*,* V *&* W*; Beacon* P*,* Q *&* R*; Monitor* E*,* F *&* G *TDIs for AMIGO-S-8-12deg* (Fig. 12)

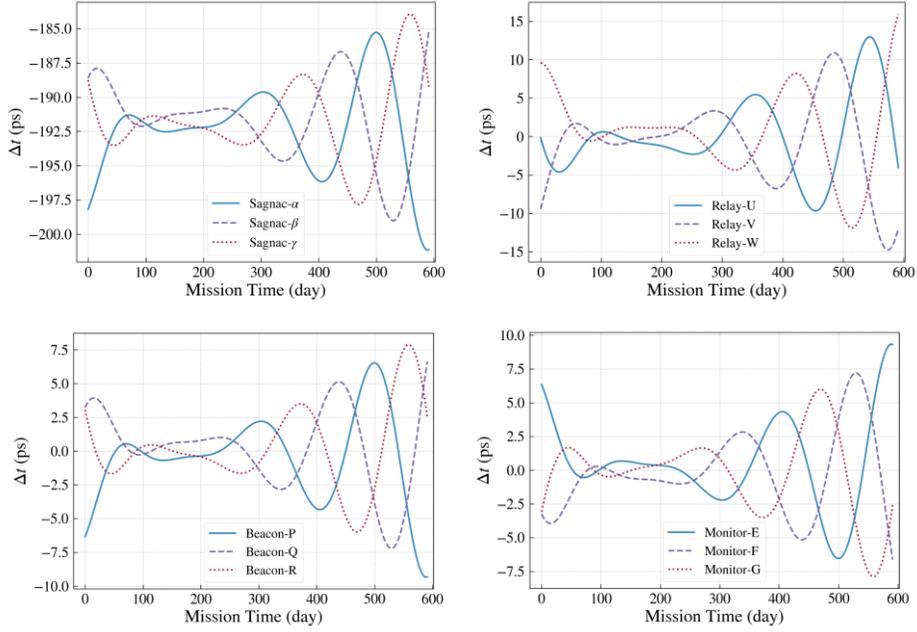

**Fig. 12.** The optical path length differences for Sagnac α, β & γ; Relay U, V & W; Beacon P, Q & R; Monitor E, F & G TDIs for AMIGO-S-8-12deg.

### 4.2. Numerical results of the first-generation TDI for AMIGO-S-2-6deg

*4.2.1. Unequal-arm Michelson X, Y & Z TDIs and their sum X+Y+Z AMIGO-EM (Fig. 13)*

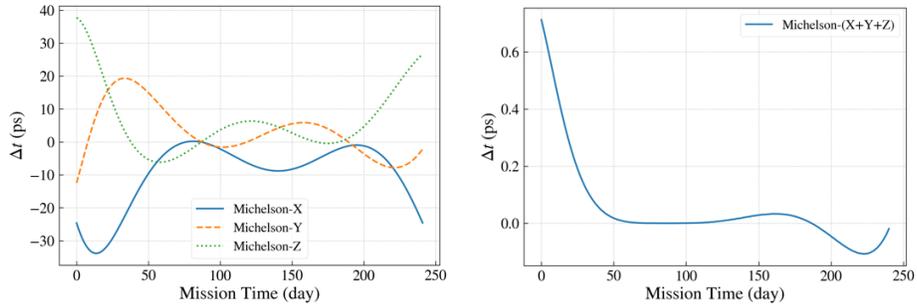

**Fig. 13.** The optical path length differences for Unequal-arm Michelson X, Y, & Z TDIs (left panel) and their sum X + Y + Z (right panel) for AMIGO-S-2-6deg.

*4.2.2. Sagnac α, β & γ; Relay U, V & W; Beacon P, Q & R; Monitor E, F & G TDIs for AMIGO-S-2-6deg* (Fig. 14)

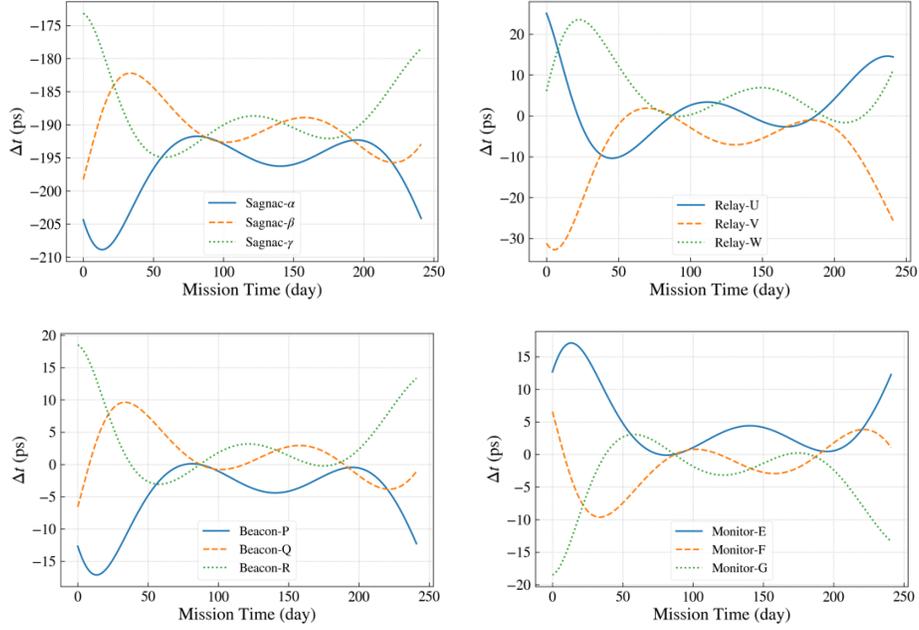

**Figure 14.** The optical path length differences for Sagnac α, β & γ; Relay U, V & W; Beacon P, Q & R; Monitor E, F & G TDIs for AMIGO-S-2-6deg.

### 4.3. Numerical results of the first-generation TDI for AMIGO-S-2-4deg

*4.3.1. Unequal-arm Michelson X, Y & Z TDIs and their sum X+Y+Z* (Fig. 15)

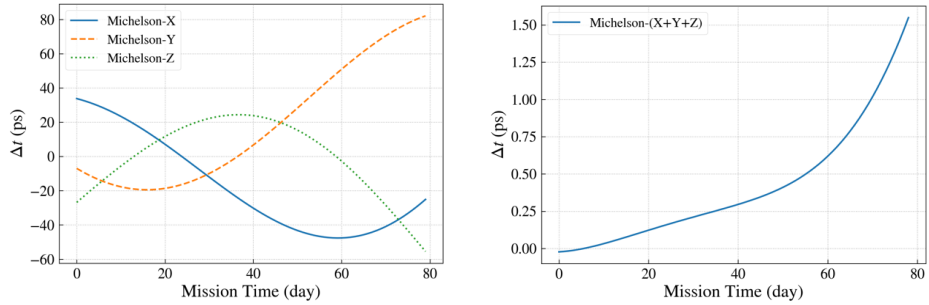

**Fig. 15.** The optical path length differences for Unequal-arm Michelson X, Y, & Z TDIs (left panel) and their sum X + Y + Z (right panel) for AMIGO-S-2deg-2030.

*4.3.2. Sagnac α, β & γ; Relay U, V & W; Beacon P, Q & R; Monitor E, F & G TDIs for AMIGO-S-2deg-2030* (Fig.16)

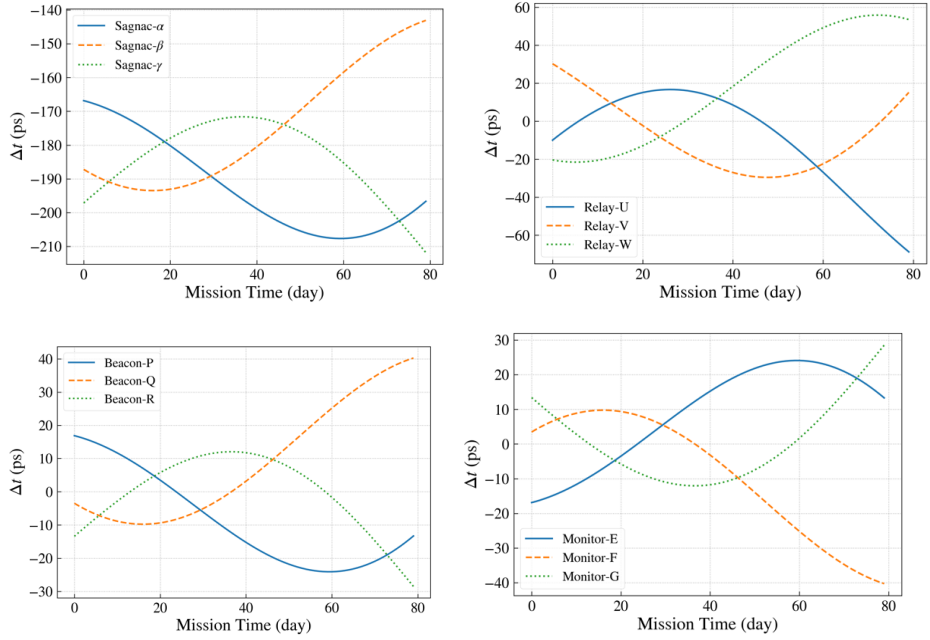

**Figure 16.** The optical path length differences for Sagnac α, β & γ; Relay U, V & W; Beacon P, Q & R; Monitor E, F & G TDIs for AMIGO-S-2-4deg.

### 4.4. Numerical results of the first-generation TDI for AMIGO-E1

*4.4.1. Unequal-arm Michelson X, Y & Z TDIs and their sum X+Y+Z AMIGO-E1* (Fig. 17)

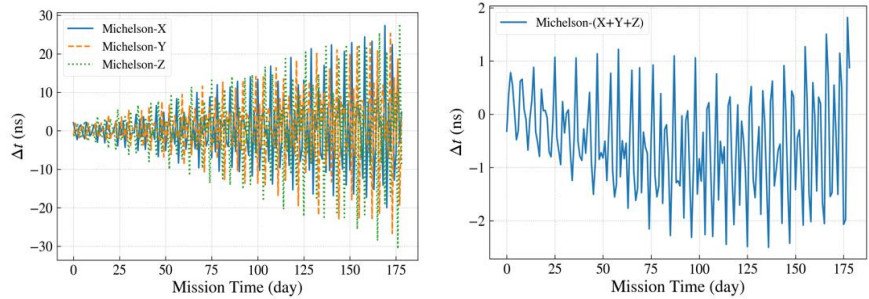

**Figure 17.** The optical path length differences for Unequal-arm Michelson X, Y, & Z TDIs (left panel) and their sum X + Y + Z (right panel) for AMIGO-E1.0.

*4.2.2. Sagnac α, β & γ; Relay U, V & W; Beacon P, Q & R; Monitor E, F & G TDIs for AMIGO-E1* (Fig. 18)

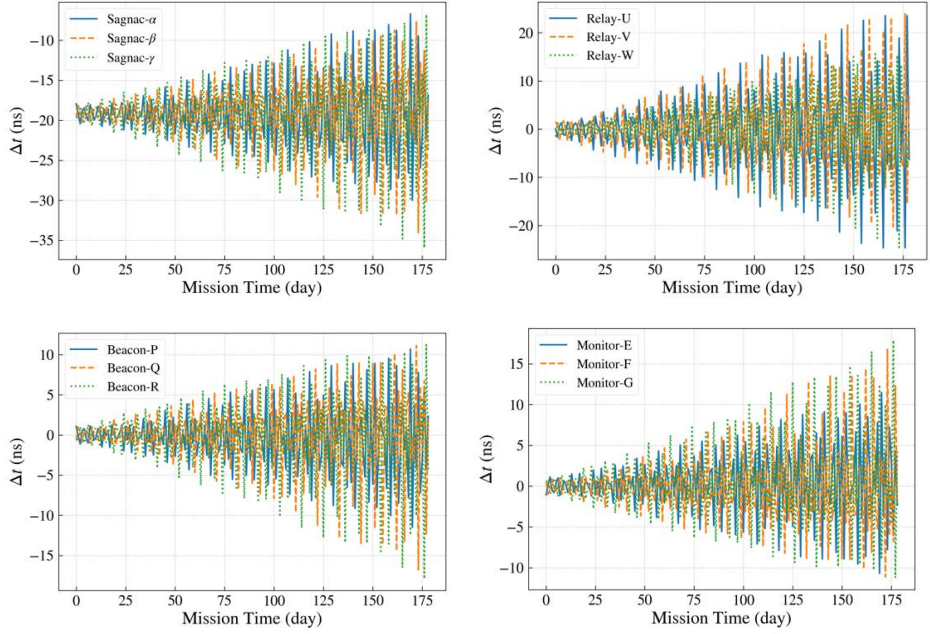

**Figure 18.** The optical path length differences for Sagnac α, β & γ; Relay U, V & W; Beacon P, Q & R; Monitor E, F & G TDIs for AMIGO-E1.

### 4.5. Numerical results of the first-generation TDI for AMIGO-EM-L4

*4.5.1. Unequal-arm Michelson X, Y & Z TDIs and their sum X+Y+Z AMIGO-EM-L4* (Fig. 19)

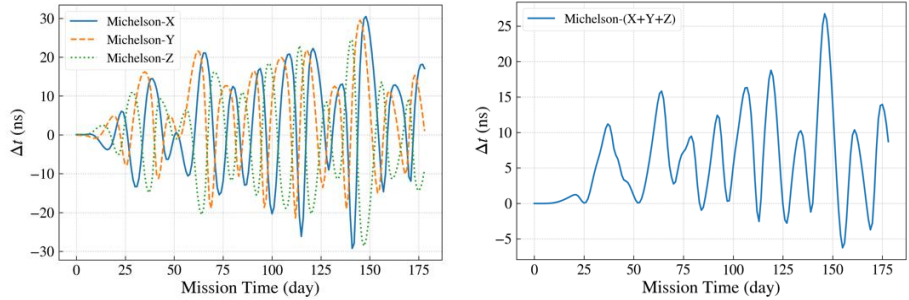

**Figure 19.** The optical path length differences for Unequal-arm Michelson X, Y, & Z TDIs (left panel) and their sum X + Y + Z (right panel) for AMIGO-EM-L4.

*4.5.2. Sagnac α, β & γ; Relay U, V & W; Beacon P, Q & R; Monitor E, F & G TDIs for AMIGO-EM-L4* (Fig. 20)

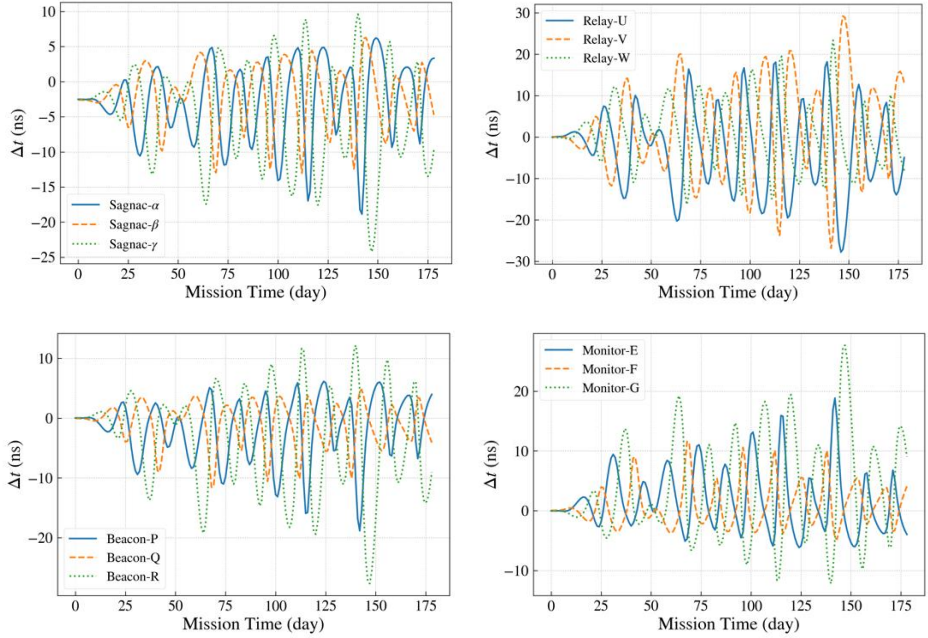

**Figure 20.** The optical path length differences for Sagnac α, β & γ; Relay U, V & W; Beacon P, Q & R; Monitor E, F & G TDIs for AMIGO-EM-L4.

## 4.6. TDI Summary (Table 5)

**Table 5.** Comparison of the resulting path length differences for the first-generation TDI's listed in [24] (i.e., X, Y, Z, X+Y+Z, Sagnac, U, V, W, P, Q, R, E, F, and G TDI configurations) for AMIGO-S (AMIGO-S-8-12deg, AMIGO-S-2-6deg, AMIGO-S-2-4deg), for AMIGO-E1 and and for AMIGO-EM, all of nominal arm lengths 10000 km.

| 1st generation TDI configuration | TDI path length difference $\Delta L$ | | | | |
|---|---|---|---|---|---|
| | AMIGO-S-8-12deg [ps] [min, max], rms average | AMIGO-S-2-6deg [ps] [min, max], rms average | AMIGO-S-2-4deg [ps] [min, max], rms average | AMIGO-E1.0 [ns] [min, max], rms average | AMIGO-EM-L4 [ns] [min, max], rms average |
| X | [-19, 14], 7 | [-34, 1], 14 | [-48, 34], 33 | [-40, 48], 15 | [-30, 31], 13 |
| Y | [-15, 14], 6 | [-13, 20], 9 | [-20, 83], 40 | [-43, 41], 16 | [-22, 30], 12 |
| Z | [-12, 16], 6 | [-7, 38], 13 | [-56, 25], 22 | [-46, 40], 16 | [-29, 25], 12 |
| X+Y+Z | [-0.01, 0.12], 0.06 | [-0.2, 0.8], 0.2 | [-0.03, 2], 0.6 | [-3, 2], 2 | [-7, 27], 9?? |
| Sagnac-α | [-202, -185], 193, 4* | [-209, -191], 197, 5 | [-208, -166], 193, 14 | [-44, -1], 21, 9 | [-19, 7], 6, 6 |
| Sagnac-β | [-200, -185], 192, 3* | [-199, -182], 191, 4 | [-194, -143], 176, 18 | [-46, -4], 20, 9 | [-14, 7], 5, 5 |
| Sagnac-γ | [-198, -183], 192, 3* | [-195, -173], 189, 6 | [-212, -171], 185, 11 | [-48, -3], 21, 9 | [-25, 10], 9, 8 |
| Relay-U | [-10, 13], 5 | [-11, 26], 8 | [-69, 17], 28 | [-42, 38], 14 | [-28, 19], 11 |
| Relay-V | [-15, 11], 6 | [-33, 2], 13 | [-30, 31], 20 | [-37, 37], 13 | [-27, 30], 12 |
| Relay-W | [-12, 16], 6 | [-2, 24], 10 | [-22, 56], 35 | [-42, 32], 13 | [-17, 24], 8 |
| Beacon-P | [-10, 7], 4 | [-18, 1], 7 | [-69, 17], 28 | [-28, 17], 8 | [-19, 7], 6 |
| Beacon-Q | [-8, 7], 3 | [-7, 10], 5 | [-30, 31], 20 | [-29, 17], 8 | [-12, 5], 4 |
| Beacon-R | [-6, 8], 3 | [-4, 19], 7 | [-22, 56], 35 | [-28, 17], 8 | [-28, 13], 10 |
| Monitor-E | [-7, 10], 4 | [-1, 18], 7 | [-17, 25], 17 | [-17, 28], 8 | [-7, 19], 6 |

| Monitor-F | [-7, 8], 3 | [-10, 7], 5 | [-41, 10], 20 | [-17, 29], 8 | [-5, 12], 4 |
| Monitor-G | [-8, 6], 3 | [-19, 4], 7 | [-13, 29], 12 | [-17, 28], 8 | [-13, 28], 10 |
| Mission duration | 600 days | 250 days | 80 days | 180 days | 180 days |
| Requirement on $\Delta L$ | 0.1 m (330 ps) | 0.1 m (330 ps) | 0.1 m (330 ps) | 0.1 m (330 ps) | 0.1 m (330 ps) |

*root mean square deviation from the mean

## 5. Thruster Requirement for Various Versions of Constant-Arm Choice of AMIGO

### 5.1. Spacecraft trajectory choices for constant-arm interferometry

After obtaining the geodesic orbits of the three S/Cs, we can identify the instantaneous plane formed by the three S/Cs by defining the following unit vectors from the instantaneous positions of the S/Cs:

$$\mathbf{n}_{23}(t) \equiv (\mathbf{r}_{S/C3} - \mathbf{r}_{S/C2}) / |\mathbf{r}_{S/C3} - \mathbf{r}_{S/C2}|,$$
$$\mathbf{n}_{21}(t) \equiv (\mathbf{r}_{S/C1} - \mathbf{r}_{S/C2}) / |\mathbf{r}_{S/C1} - \mathbf{r}_{S/C2}|,$$
$$\mathbf{n}_z(t) \equiv (\mathbf{n}_{23} \times \mathbf{n}_{21}) / |\mathbf{n}_{23} \times \mathbf{n}_{21}|, \quad (5)$$

where $\mathbf{r}_{S/Ci}$ is the instantaneous positions of the S/C$i$ ($i$ = 1, 2, 3) at time $t$. From Eq. (5), we define two orthogonal unit vectors in the instantaneous plane of the S/C formation:

$$\mathbf{n}_1 \equiv \mathbf{n}_{23},$$
$$\mathbf{n}_2 \equiv \mathbf{n}_z \times \mathbf{n}_1. \quad (6)$$

Without loss of generality, we let S/C2 and S/C3 follow the geodesic trajectory of S/C1 as follows:

$$\mathbf{r}_{traj,S/C1} = \mathbf{r}_{S/C1},$$
$$\mathbf{r}_{traj,S/C2} = \mathbf{r}_{S/C1} - (l/2) \mathbf{n}_1 - (3^{1/2}/2) \mathbf{n}_2,$$
$$\mathbf{r}_{traj,S/C3} = \mathbf{r}_{S/C1} + (l/2) \mathbf{n}_1 - (3^{1/2}/2) \mathbf{n}_2, \quad (7)$$

where $\mathbf{r}_{traj,S/C1}$, $\mathbf{r}_{traj,S/C2}$ and $\mathbf{r}_{traj,S/C3}$ are the aimed trajectories of S/C1, S/C2 and S/C3 respectively, and $l$ is the nominal arm length 10000 km.

### 5.2. Thruster acceleration

From Eq. (7), the acceleration at a specific point in a trajectory is calculated as the second derivative of position with respect to time,

$$\mathbf{a}_{traj} = d^2\mathbf{r}_{traj}/dt^2. \quad (8)$$

At this specific point, the gravitational (free-fall) acceleration $\mathbf{a}_{eph}$ in the ephemeris is given by:

$$\mathbf{a}_{\text{traj}}(\mathbf{r}_{\text{traj}}, d\mathbf{r}_{\text{traj}}/dt) = \mathbf{a}_{\text{Newton}} + \mathbf{a}_{\text{1PN}} + \mathbf{a}_{\text{fig}} + \mathbf{a}_{\text{asteroid}}, \tag{9}$$

where $\mathbf{a}_{\text{Newton}}$ and $\mathbf{a}_{\text{1PN}}$ are the Newtonian and first-order post-Newtonian acceleration from the major celestial bodies in the solar system considered as point mass, $\mathbf{a}_{\text{fig}}$ is the acceleration due to the figure effects from the Sun, Earth and Moon, and $\mathbf{a}_{\text{asteroid}}$ is the acceleration from Newtonian perturbation of the 340 asteroids. The explicit interactions in our CGC ephemeris framework are fully described in references [24, 36, 37].

The thruster needs to provide acceleration $\mathbf{a}_{\text{thruster}}$

$$\mathbf{a}_{\text{thruster}} = \mathbf{a}_{\text{traj}} - \mathbf{a}_{\text{traj}} \tag{10}$$

to maintain the constant arm length trajectories.

For calculation of $a_{\text{thruster}}$, we choose the three AMIGO orbit configurations AMIGO-E1 around the Earth (Section 3.1), and AMIGO-EML4 near the Earth-Moon L4 point. The corresponding accelerations needed vs time are shown in Figure 21.

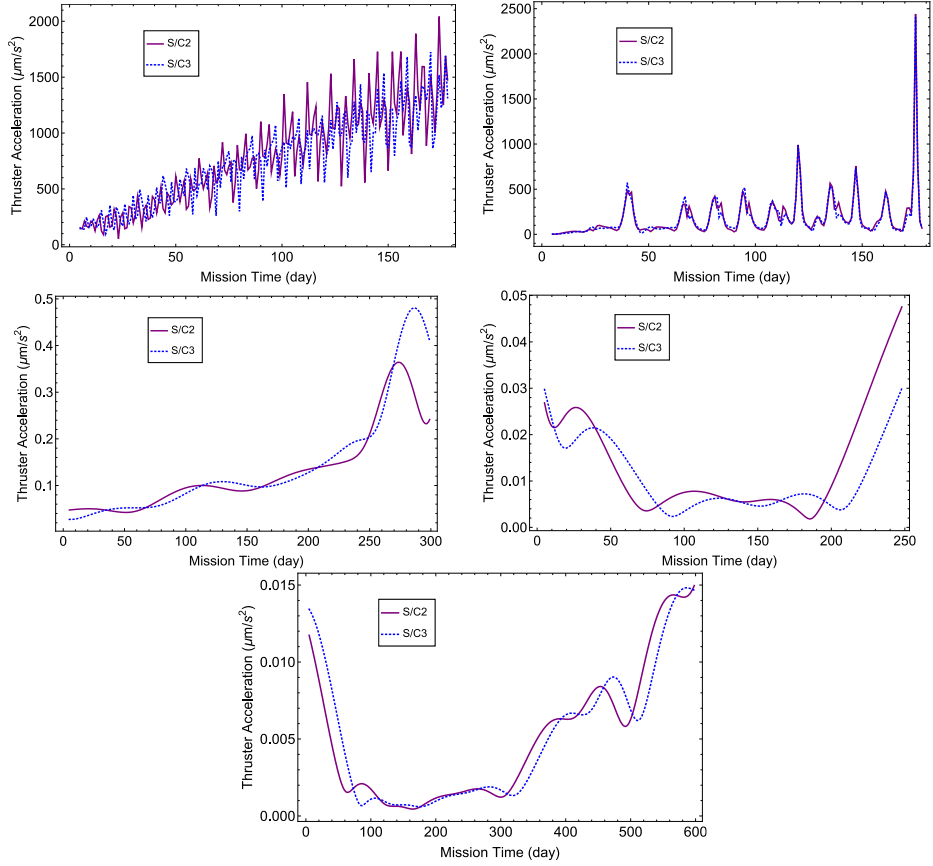

**Fig. 21**. The thruster acceleration compensations to maintain the 10,000 km arm length for S/Cs in the ephemeris framework CGC3.0 for AMIGO-E1 (upper left panel), AMIGO-EML4 (upper right panel), AMIGO-S (2-4deg behind the Earth) (middle right panel), AMIGO-S (2-6deg behind the Earth) (middle right panel) and AMIGO-S (8-12deg behind the Earth) (lower panel) configurations, respectively.

### 5.3. Fuel requirement (Table 6)

**Table 6**. The thruster and propellant requirement for AMIGO mission assuming the mass of the S/C is 1000 kg.

| Mission Concept (arm length $10^4$ km) | Required acceleration (max) | Thruster requirement (max) | Propellant requirement for 1 yr by numerical integration (kg) | |
|---|---|---|---|---|
| | | | $I_{sp} = 300$ sec | $I_{sp} = 1000$ s |
| AMIGO-E1 | 2.0 mm/s$^2$ | 2.0 N | 999.8 | 922.0 |
| AMIGO-EML4 | 2.5 mm/s$^2$ | 2.5 N | 863.0 | 449.2 |
| AMIGO-S-2-4-deg | 500 nm/s$^2$ | 500 μN | 1.54 | 0.464 |
| AMIGO-S-2-6-deg | 50 nm/s$^2$ | 50 μN | 0.15 | 0.045 |
| AMIGO-S-8-12-deg | 15 nm/s$^2$ | 15 μN | 0.05 | 0.016 |

Note in the AMIGO-S-2-4-deg case, we have extended the 80-day design of to 300 days for this constant equal-arm option. This tells us the operational period could be extended further compared to geodetic options.

### 6. Discussion and Outlook

Since middle frequency band (0.1-10 Hz) is an important for GW detection to bridge the gap between high-frequency GW detection and low-frequency GW detection, we have proposed AMIGO for this purpose [6]. On Earth, there are severe gravity gradient fluctuations in the middle frequency band. These fluctuation needs to be measured/estimated before the sensitivity of GW detection on Earth can be increased [11]. Space seems to be a good place to do GW detection in the middle frequency band in addition to low frequency band (100 nHz-100 mHz). The situation is especially so after the success of drag-free demonstration of LISA Pathfinder [21, 22].

In this paper, we present the mission concept with detailed mission orbit studies, TDIs and constant equal arm options of AMIGO. For geocentric orbit options: (i) in the geodetic schemes, the formation angles deviate from 60° rather quickly and the first-generation TDIs do not satisfy the requirement; (ii) in the constant equal arm schemes, the required accelerations are 3-5 orders larger than those of heliocentric orbit options and the propellant requirements are untenable. For heliocentric options: (i) in the geodetic schemes, the formation angles deviate mildly from 60° and the first-generation TDIs requirements are well satisfied; (ii) in the constant equal arm schemes, the required accelerations are in the 15-500 nm/s$^2$ range and the propellant & thruster requirements are tenable. *Therefore, the heliocentric orbit options would be the first choice.*

As to the choice between the geodetic schemes and the constant equal arm schemes: (i) for the geodetic schemes, although the requirement on the first-generation TDI is well satisfied by the orbit, the local requirement on the amplification noise, the timing noise and the clock noise etc. are still 0.1 m which is much stringent than the longer arm space mission. Although 0.1 m accuracy is feasible and implementable with a more stable clock on board, it is much more demanding. (ii) for the constant equal arm schemes, although the required accelerations are in the nm/s$^2$ range and the propellant & thruster requirements are feasible and implementable, the inertial sensor/accelerometer need to be actuated. However, the gap size of the current inertial sensor/accelerometer limits the total range of one acceleration maneuver to about 2 mm. This limits the one acceleration maneuver time of AMIGO-S to about 500 sec. A reference is needed for the measurement/monitoring of the actuation. Hence an alternate proof mass is needed. Laser metrology has the required accuracy. The two proof masses can alternate to become the reference masses. This way the required dynamical range can be achieved. The actuation induced Fourier spectral components needs to be subtracted. These technic issues seem tenable and need to be tackled.

Constant equal-arm Michelson interferometry is preferred if the technic issues can be resolved, because it does not have the complication of the TDI. Moreover, constant equal-arm Michelson interferometry and the TDI could both be tested at the beginning of science mission and worked out in the same mission if pre-mission preparation was done.

DECIGO and B-DECIGO are constant equal-arm GW missions having nominal arm length 1000 km and 100 km. The constant arm-length orbit formations of AMIGO could be rescaled to these lengths to become DECIGO and B-DECIGO orbit options [44-46]. It could also rescale to the recently proposed TianGO GW mission [47] whether it adopted constant arm or not. In accompanying papers, we also study the orbit formations for AIGSO [48, 49] and the constant equal-arm orbit options for LISA and TAIJI [50].

The AMIGO-S (8°-12° behind Earth) orbits starting at Epoch JD2462316.0 (2029-Jun-28th 12:00:00) for 600 days could be an earlier geodetic GW mission option. If a 10 year geodetic mission is desired, it has to go to about 20° behind the Earth orbit. The AMIGO-S (2°-6° behind Earth) orbits starting at Epoch JD2462416 (2029-Oct-6th 12:00:00) for 250 days and the AMIGO-S (2°-4° behind Earth) orbits starting at Epoch JD2462503.0 (2030-Jan-1st 12:00:00) (for 80 days in the geodetic option; for 300 days or more for the constant equal-arm option) could be a pathfinder mission; they are closer to Earth and takes less days and less power for deployments.

## Acknowledgements

This work was supported by Strategic Priority Research Program of the Chinese Academy of Sciences under grant Nos. XDA and XDB21010100, and National Key Research and Development Program of China under Grant Nos. 2016YFA0302002 and 2017YFC0601602.